\documentclass[journal]{IEEEtran}

\usepackage{amsmath,amssymb,amsfonts}
\usepackage{mathtools}
\usepackage{graphicx}
\usepackage{cite}
\usepackage[caption=false,font=footnotesize]{subfig}
\usepackage{balance}
\usepackage{xcolor}
\usepackage[hidelinks,colorlinks=true,linkcolor=blue,
            citecolor=blue,urlcolor=black]{hyperref}

\begin{document}

\title{Dual-Pilot Design for OCDM Channel Estimation}

\author{Ayoub Ammar Boudjelal
and H\"{u}seyin Arslan,~\IEEEmembership{Fellow,~IEEE}
\thanks{The authors are with the Department of Electrical and
Electronics Engineering, Istanbul Medipol University, Istanbul,
34810, Turkey (e-mail:
ayoub.ammar@std.medipol.edu.tr;
huseyinarslan@medipol.edu.tr).}
}

\AddToHook{shipout/foreground}{\put(0.13\paperwidth,-0.035\paperheight){\parbox[t]{0.74\paperwidth}{\footnotesize\centering This work has been accepted for publication in IEEE Wireless Communications Letters. DOI: 10.1109/LWC.2026.3756907}}}
\AddToHookNext{shipout/foreground}{\put(0.1\paperwidth,-0.955\paperheight){\parbox[t]{0.8\paperwidth}{\scriptsize\centering \copyright~2026 IEEE. Personal use of this material is permitted. Permission from IEEE must be obtained for all other uses, in any current or future media, including reprinting/republishing this material for advertising or promotional purposes, creating new collective works, for resale or redistribution to servers or lists, or reuse of any copyrighted component of this work in other works.}}}
\maketitle

\begin{abstract}
Channel acquisition in orthogonal chirp division multiplexing (OCDM)
over doubly-dispersive channels incurs a large pilot overhead, since
delay and Doppler jointly shift the symbols in the chirp domain and
separating them requires repeated pilots, each protected against the
complete coupled displacement. This letter proposes a dual-pulsone
pilot design that reduces this overhead in two steps. First, a pulsone
occupies fewer chirp-domain bins than alias-free Doppler estimation
requires, deliberately allowing Doppler wrapping. Second, a
phase-offset pulsone is placed within the same protected region to
observe a complementary part of the displacement, further reducing
the shared guard. The phase offset also resolves the wrapping
ambiguity, enabling low-complexity delay-Doppler estimation.
Simulations show competitive estimation and BER performance, with the
lowest complexity and highest payload efficiency among the considered
OCDM estimators.
\end{abstract}

\begin{IEEEkeywords}
Orthogonal chirp division multiplexing (OCDM), channel estimation,
high-mobility communications, fractional Doppler, pulsone.
\end{IEEEkeywords}

\IEEEpeerreviewmaketitle

\section{Introduction}
\label{sec:introduction}

\IEEEPARstart{H}{igh-mobility} wireless channels exhibit simultaneous
delay and Doppler dispersion, resulting in severe time and frequency
selectivity~\cite{10742574,10382693}. Orthogonal chirp division
multiplexing (OCDM) is attractive in such environments because the
discrete Fresnel transform (DFnT) spreads every symbol over the
available time-frequency resources while retaining a close
implementation relationship with OFDM~\cite{7523229,9154217}, provided
that the channel is acquired without excessive pilot overhead or
receiver complexity.

Earlier OCDM estimators mainly addressed static or frequency-selective
channels~\cite{9597526,9997134,9424142} or a single carrier-frequency
offset~\cite{9638964}. For doubly-selective channels, pilot
chirp-assisted OCDM (PCA-OCDM)~\cite{wang2023pilot} optimizes pilot
chirps for basis-expansion-model (BEM) MMSE estimation; delay-Doppler
based OCDM (DD-OCDM)~\cite{zegrar2024novel} designs Fresnel-domain
pilots and estimates the channel in the delay-Doppler domain; DFT-OCDM
\cite{haif2024novel} adds DFT processing to remove the delay-Doppler
coupling and uses time-domain windowing against fractional-Doppler
leakage; and orthogonal matching pursuit OCDM
(OMP-OCDM)~\cite{zhan2024omp} superimposes pilots and data and applies
iterative sparse recovery. In orthogonal time-frequency space (OTFS),
guard-free superimposed pilots have also been combined with
variational-inference-based channel-coefficient
estimation~\cite{wu2025isac}, at the cost of iterative pilot-data
interference handling. These approaches improve channel acquisition but
add cost through statistical estimation, full-domain transformations,
modified transceivers, or iterative processing.

The underlying difficulty is that delay and Doppler jointly
determine the chirp-domain displacement of each
path~\cite{haif2024novel}, while Doppler also induces a phase rotation
across the chirp index. Separating them therefore requires repeated
pilots, each protected against the complete coupled displacement, so
the overhead grows with both the number of pilot bins and the guard
size. In contrast, OTFS and affine frequency division multiplexing
(AFDM) separate delay and Doppler, allowing the guard to match the
channel support~\cite{raviteja2019otfs,bemani2023afdm}. Existing two-pilot designs, such as DFT-OCDM, which
recovers delay from the phase relation of two
pilots~\cite{haif2024novel}, use the additional pilot for further
observations, while the protected region must accommodate the complete
channel spread. In contrast, this letter proposes dual-pulsone OCDM
(DP-OCDM), which reduces the overhead in two steps. First, a pulsone
occupies fewer chirp-domain bins than alias-free Doppler estimation
requires, deliberately allowing Doppler wrapping. Second, a
phase-offset pulsone is placed within the same protected region to
observe a complementary part of the displacement, further reducing the
shared guard. The phase offset also maps the two responses to
distinguishable Doppler coordinates, resolving pilot association and
the wrapping ambiguity. Together, the two pulsones enable
low-complexity channel estimation using a few localized short
transforms.

The main contributions are:
\begin{itemize}
\setlength{\itemsep}{1pt}
\setlength{\parskip}{0pt}

    \item A dual-pulsone pilot design that reduces the
    channel-acquisition overhead in two ways: the phase offset between
    the pulsones resolves pilot association under controlled Doppler
    wrapping, reducing the number of bins occupied by each pulsone,
    while the second pulsone observes a complementary displacement
    interval within the same protected region, shrinking the guard at
    every pilot bin by the maximum Doppler spread.

    \item The periodic dual-pulsone structure is exploited
    for low-complexity channel estimation, in which delay, Doppler, and
    gain are recovered from a few localized short transforms with
    guaranteed uniqueness under Doppler wrapping, and fractional
    Doppler is handled by a closed-form refinement without transceiver
    modification.

    \item Simulations show competitive NMSE and BER, the
    lowest channel-acquisition complexity, and the highest payload
    efficiency among the considered OCDM estimators.
\end{itemize}

\begin{figure*}[t!]
    \centering
    \includegraphics[width=0.96\textwidth]{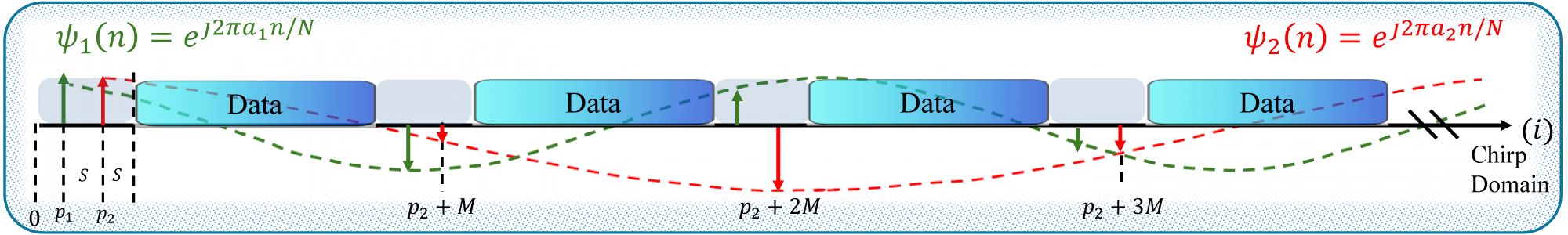}
    \caption{Proposed DP-OCDM frame.}
    \label{fig:pulsone_pilot}
\end{figure*}

\section{OCDM Signal Model and Pilot-Guard Bottleneck}
\label{sec:model}

Let $L$ denote the OCDM block length and $x(i)$,
$i=0,\ldots,L-1$, the chirp-domain symbols. The time-domain block is
generated by the inverse DFnT as~\cite{7523229}
$$
s(p)=
\frac{e^{-\jmath\pi/4}}{\sqrt L}
\sum_{i=0}^{L-1}
x(i)e^{-\jmath\pi(p-i)^2/L}.
$$
A CP of length $N_{\rm CP}\geq\ell_{\max}$ is assumed. Throughout,
$[a]_L$ denotes reduction of $a$ modulo $L$.

Consider an $R$-path integer-grid channel. Path $r$ has effective
complex gain $h_r$, delay $0\leq\ell_r\leq\ell_{\max}$, and
normalized Doppler $-\nu_{\max}\leq\nu_r\leq\nu_{\max}$. If $T_s$
is the sampling interval, its physical Doppler is
$f_{D,r}=\nu_r/(LT_s)$. After CP removal and DFnT demodulation, with
the deterministic Fresnel phase absorbed into $h_r$, the received
chirp-domain signal is~\cite{haif2024novel}
\begin{equation}
Y_c(i)=
\sum_{r=1}^{R}
h_r
e^{\jmath\frac{2\pi}{L}\nu_r(i-\ell_r-\nu_r)}
x([i-\ell_r-\nu_r]_L)+w(i),
\label{eq:chirp_io}
\end{equation}
where $w(i)\sim\mathcal{CN}(0,\sigma^2)$. We assume that delays are integers, and that $\ell_{\max}$ and $\nu_{\max}$
are known for pilot design. Doppler shifts are integers except in
Section~\ref{sec:fractional}, where a fractional offset is added under
the separation condition stated there.

Define the channel-induced displacement
$\Delta_r\triangleq\ell_r+\nu_r$ and
$S\triangleq\ell_{\max}+\nu_{\max}$. Thus,
$-\nu_{\max}\leq\Delta_r\leq S$. The resulting relative pilot-data
displacement satisfies
$$
-S-\nu_{\max}
\leq
\Delta_1-\Delta_2
\leq
S+\nu_{\max}.
$$
Hence, an isolated chirp-domain pilot requires
$2(S+\nu_{\max})+1$ protected positions and $2\nu_{\max}$ pilot
repetitions, resulting in considerable overhead.

\section{Dual-Pulsone Pilot Design}
\label{sec:frame}

Factor the OCDM block as $L=MN$ and write $i=i_g+nM$, where
$0\leq i_g<M$ and $0\leq n<N$. The design uses
$N\geq\max\{\nu_{\max}+1,\ell_{\max}+1\}$, with $N$ chosen as a
divisor of $L$. Unlike an alias-free Doppler representation requiring
$N\geq2\nu_{\max}+1$, this permits a shorter periodic dimension.

The two pulsones are placed at $p_1=\nu_{\max}$ and
$p_2=\nu_{\max}+S+1$, with phase indices satisfying
$[a_2-a_1]_N=\nu_{\max}$. Accordingly, let $a_1=\alpha$ and
$a_2=[\alpha+\nu_{\max}]_N$, where
$\alpha\in\{0,\ldots,N-1\}$. We set $\alpha=0$ hereafter for
simplicity. Define
$K\triangleq\nu_{\max}+2S+2
=3\nu_{\max}+2\ell_{\max}+2$ and require $M>K$.
For $k\in\{1,2\}$, the pulsone sequence is
$$
\psi_k(n)=
\sqrt{\frac{E_p}{N}}\,
e^{\jmath2\pi a_kn/N},
\qquad n=0,\ldots,N-1,
$$
where $E_p$ is the per-pulsone energy. The chirp-domain frame
\begin{equation}
x(i_g+nM)=
\begin{cases}
\psi_k(n), & i_g=p_k,\\
0, & 0\leq i_g<K,~i_g\notin\{p_1,p_2\},\\
\sqrt{E_d}\,d(i_g,n), & K\leq i_g<M,
\end{cases}
\label{eq:frame}
\end{equation}
where $\mathbb E[|d(i_g,n)|^2]=1$ and $E_d$ is the average
data-symbol energy. Fig.~\ref{fig:pulsone_pilot} illustrates the
resulting DP-OCDM frame, whose channel-acquisition overhead is
$\varphi_{\rm DP}=NK$.

To determine the useful observation region, note that data begins at
$i_g=K$. Its largest negative displacement reaches
$K-\nu_{\max}$, while positively shifted data wrapping from the
preceding residue period can occupy only $i_g=0,\ldots,S-1$.
Therefore, $i_g=S,\ldots,2S+1$ are free of shifted data.

The boundary displacement $\Delta_r=\ell_{\max}$ appears through
$\psi_1$ at $i_g=S$ and through $\psi_2$ at $i_g=2S+1$. To avoid
duplicate detection, it is assigned to $\psi_2$, giving
\begin{equation}
S+1\leq i_g\leq2S+1.
\label{eq:search_region}
\end{equation}
Within this region, $\psi_1$ observes
$\ell_{\max}+1\leq\Delta_r\leq S$, whereas $\psi_2$ observes
$-\nu_{\max}\leq\Delta_r\leq\ell_{\max}$. The intervals are
disjoint and jointly cover the complete admissible displacement
support. Thus, the two pulsones share one protected region instead of
independently protecting the complete channel shift.

The reduction results from two separable
choices. A single pulsone needs $2S+2\nu_{\max}+1$ protected positions
per period, i.e.,
$\varphi_{\rm 1P}^{\rm af}=N_{\rm af}(2S+2\nu_{\max}+1)$ with an
alias-free dimension $N_{\rm af}\geq2\nu_{\max}+1$. Modulo-$N$
wrapping alone reduces this by the factor $N/N_{\rm af}$, since one
pulsone remains identifiable for $N>\ell_{\max}$
(Section~\ref{sec:estimation}). The complementary construction then
yields $K=2S+\nu_{\max}+2$, saving $\nu_{\max}-1$ further positions
per period.

\section{Localized Delay-Doppler Estimation}
\label{sec:estimation}

For each residue in \eqref{eq:search_region}, the receiver collects
$Y_c(i_g+nM)$, $n=0,\ldots,N-1$, and applies the unitary
$N$-point IFFT for $m=0,\ldots,N-1$,
\begin{equation}
Z(m,i_g)=
\frac{1}{\sqrt N}
\sum_{n=0}^{N-1}
Y_c(i_g+nM)e^{\jmath2\pi nm/N}.
\label{eq:localized_ifft}
\end{equation}

For a path observed through $\psi_1$,
$i_g=p_1+\ell_r+\nu_r$, and for $\psi_2$,
$i_g=p_2+\ell_r+\nu_r$. Substituting
\eqref{eq:frame} and \eqref{eq:chirp_io} into
\eqref{eq:localized_ifft} gives
\begin{equation}
\begin{aligned}
Z(m,p_1+\ell_r+\nu_r)
&=
\frac{\sqrt{E_p}h_r}{N}
e^{\jmath2\pi\nu_rp_1/L}
\\[-1mm]
&\quad\times
\sum_{n=0}^{N-1}
e^{\jmath2\pi n(m+\nu_r)/N}
+W_z,
\\[1mm]
Z(m,p_2+\ell_r+\nu_r)
&=
\frac{\sqrt{E_p}h_r}{N}
e^{\jmath2\pi\nu_rp_2/L}
\\[-1mm]
&\quad\times
\sum_{n=0}^{N-1}
e^{\jmath2\pi n(m+\nu_{\max}+\nu_r)/N}
+W_z .
\end{aligned}
\label{eq:pulsone_responses}
\end{equation}
Here $W_z\sim\mathcal{CN}(0,\sigma^2)$.

Defining $\mu\triangleq[-m]_N$, the peak coordinates are
$$
\begin{aligned}
\psi_1:\quad&
(\mu,i_g)=
([\nu_r]_N,p_1+\ell_r+\nu_r),
\\
\psi_2:\quad&
(\mu,i_g)=
([\nu_{\max}+\nu_r]_N,p_2+\ell_r+\nu_r).
\end{aligned}
$$
\begin{figure}[t!]
\centering
\includegraphics[width=0.47\textwidth]{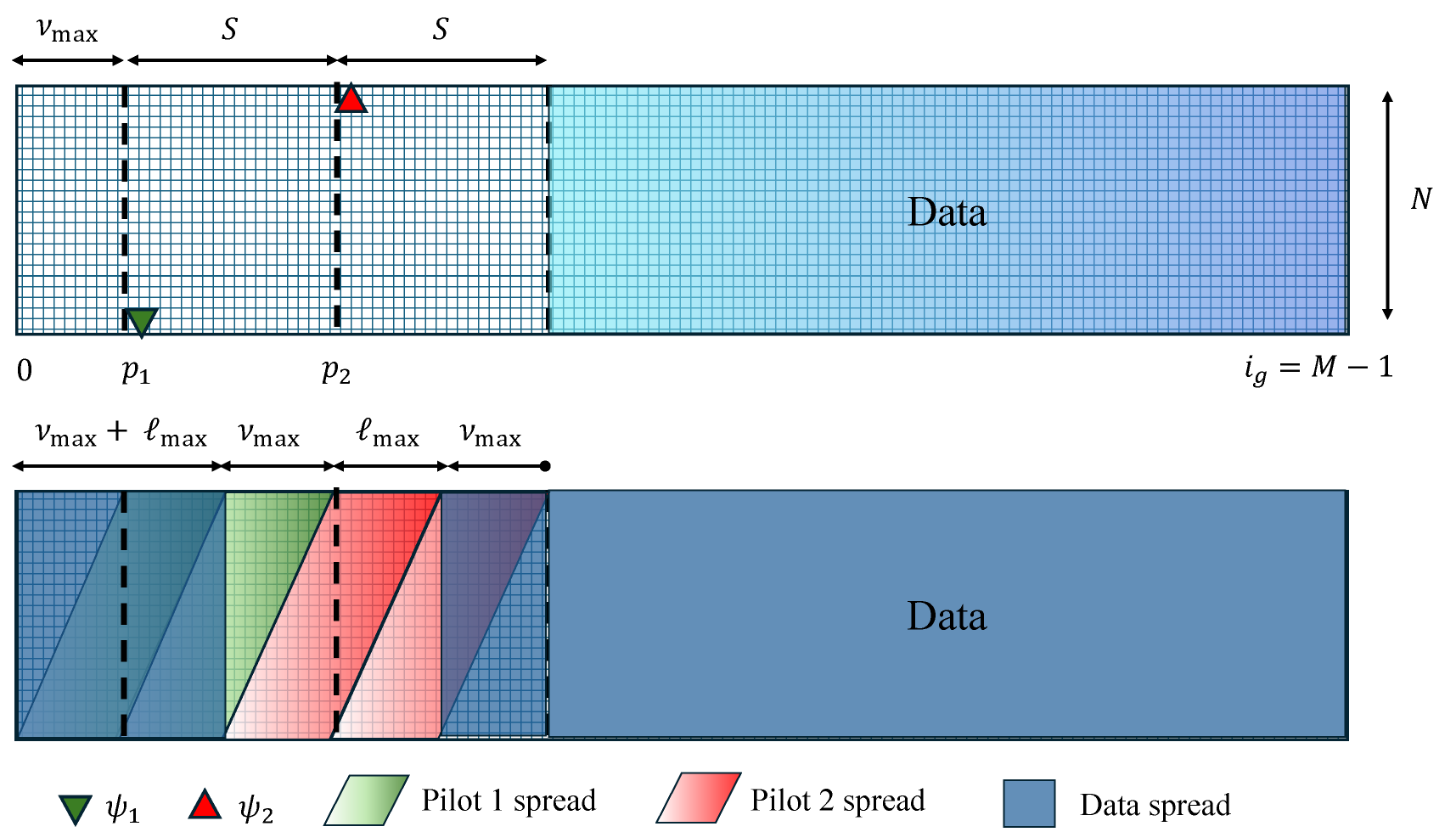}
\caption{Localized-transform observation of DP-OCDM. $\psi_1$ and
$\psi_2$ map complementary displacement intervals into the
data-free search region.}
\label{fig:DP}
\end{figure}
Thus, $\mu$ reveals Doppler modulo $N$, while $i_g$ reveals the
coupled displacement $\ell_r+\nu_r$. Fig.~\ref{fig:DP} illustrates
the resulting complementary pulsone responses and the common
data-free observation region.

For threshold-based support detection, the search contains
$N_{\rm s}=(S+1)N$ bins. On the integer grid, this region
is data-free and the unitary IFFT keeps noise-only bins i.i.d.\
$\mathcal{CN}(0,\sigma^2)$. Thus, for frame-level false-alarm
probability $P_{\rm FA}$, the threshold is
\begin{equation}
\gamma=
\sigma
\sqrt{
-\ln\!\left[
1-(1-P_{\rm FA})^{1/N_{\rm s}}
\right]
}.
\label{eq:detection_threshold}
\end{equation}
Bins satisfying $|Z(m,i_g)|>\gamma$ are retained as path candidates.

For a detected pair $(\hat m,\hat i_g)$, let
$\hat\mu=[-\hat m]_N$. The two hypotheses are
\begin{equation}
\begin{aligned}
\psi_1:\quad&
\hat\nu=\hat\mu+qN,
&
\hat\ell=\hat i_g-p_1-\hat\nu,
\\
\psi_2:\quad&
\hat\nu=\hat\mu-\nu_{\max}+qN,
&
\hat\ell=\hat i_g-p_2-\hat\nu,
\end{aligned}
\label{eq:channel_recovery}
\end{equation}
where $q\in\mathbb Z$ is chosen so
$|\hat\nu|\leq\nu_{\max}$ and
$0\leq\hat\ell\leq\ell_{\max}$.

The estimate remains unique even for $N<2\nu_{\max}+1$. For the same
pulsone, a collision requires
$\nu_1-\nu_2=qN$ and $\ell_1-\ell_2=-qN$. Since
$\ell_{\max}<N$, only $q=0$ is admissible. A cross-pulsone collision
requires
$$
\nu_1-\nu_2=\nu_{\max}+qN,
\qquad
\ell_1-\ell_2=\ell_{\max}+1-qN.
$$
For $q\leq0$, the delay constraint is violated, whereas for $q\geq1$,
$|\nu_1-\nu_2|\geq\nu_{\max}+N>2\nu_{\max}$, violating the Doppler
support. Hence, each admissible peak corresponds to a unique physical
delay-Doppler pair.

Once the pulsone index $k$ is identified, the effective path gain is
$$
\hat h_r=
\frac{Z(\hat m_r,\hat i_{g,r})}{\sqrt{E_p}}
e^{-\jmath2\pi\hat\nu_rp_k/L},
$$
and, conditioned on correct detection,
$\mathbb E[|\hat h_r-h_r|^2]=\sigma^2/E_p$, since distinct
integer-grid paths occupy distinct bins.

For power-of-two $N$, since only $S+1$ residues are transformed, the
split-radix count \cite{zegrar2024novel} is approximately
$$
C_{\rm DP}
\simeq
\frac{(S+1)N}{3}\log_2N
$$
complex multiplications. Thresholding and scalar parameter recovery
are negligible in comparison.


\section{Fractional-Doppler Estimation}
\label{sec:fractional}

We next extend the estimator to fractional Doppler while retaining the
integer-delay assumption. Write
$\widetilde{\nu}_r=\nu_r+\epsilon_r$, where
$-1/2<\epsilon_r\leq1/2$. For integer Doppler, a path observed through
pulsone $k$ has coordinates $\mu_r=[a_k+\nu_r]_N$ and
$i_{g,r}=p_k+\ell_r+\nu_r$. Replacing $\nu_r$ by
$\widetilde{\nu}_r$ spreads the response around these coordinates.
Since $\mu=[-m]_N$, the fractional displacement is oriented along
$(\Delta m,\Delta i_g)=(-\epsilon_r,\epsilon_r)$.

Let $(m_r^{(0)},i_{g,r}^{(0)})$ be the coarse integer-grid peak and
$\hat\nu_{0,r}$ the corresponding integer Doppler estimate. Samples
along the relevant diagonal are
\begin{equation}
z_r(u)=
Z\!\left([m_r^{(0)}-u]_N,i_{g,r}^{(0)}+u\right),
\qquad u\in\mathbb Z ,
\label{eq:frac_diag_samples}
\end{equation}
where $u=0$ denotes the coarse peak.

Using $L=MN$ and the periodic pulsone structure, define the
normalized Dirichlet kernel
$$
D_J(x)=
\frac{1}{J}\sum_{q=0}^{J-1}e^{\jmath2\pi qx/J}.
$$
For $x=\epsilon_r-u$, the $N$ periodically spaced replicas then satisfy
$$
\sum_{v=0}^{N-1}
e^{\jmath2\pi(a_k-p_k)v/N}
D_L(x+vM)
=
e^{\jmath2\pi[p_k-a_k]_N x/L}D_M(x).
$$
Hence, after the localized transform, the fractional response along the
diagonal becomes
\begin{equation}
z_r(u)=
h_r\sqrt{E_p}\,
e^{\jmath\phi_{k,r}(u)}
D_M(\epsilon_r-u)+w_r(u),
\label{eq:frac_diag_response}
\end{equation}
where $w_r(u)$ includes noise and residual interference,
$$
D_M(x)=
e^{\jmath\pi(M-1)x/M}
\frac{\sin(\pi x)}{M\sin(\pi x/M)},
$$
and
\begin{align}
\phi_{k,r}(u)
=&\,
\frac{\pi}{L}
\left[
(i_{g,r}^{(0)}+u-\ell_r)^2-p_k^2
-\widetilde{\nu}_r^{\,2}
\right]
\nonumber\\[-1mm]
&+
\frac{2\pi[p_k-a_k]_N}{L}
(\epsilon_r-u).
\label{eq:frac_phase}
\end{align}
For $\epsilon_r=0$, the response is concentrated at $u=0$, recovering
the integer-grid result. The refinement assumes that the used samples $z_r(u)$ are
dominated by the detected path, i.e., the responses of different paths
are sufficiently separated in the localized domain. Residual leakage
from other paths and data is then absorbed in $w_r(u)$, so
\eqref{eq:detection_threshold} becomes approximate; strongly
overlapping fractional responses may bias the estimates and are not
treated here.

\begin{figure*}[t!]
\centering
\subfloat[NMSE.\label{fig:nmse}]
{\includegraphics[width=0.23\textwidth]{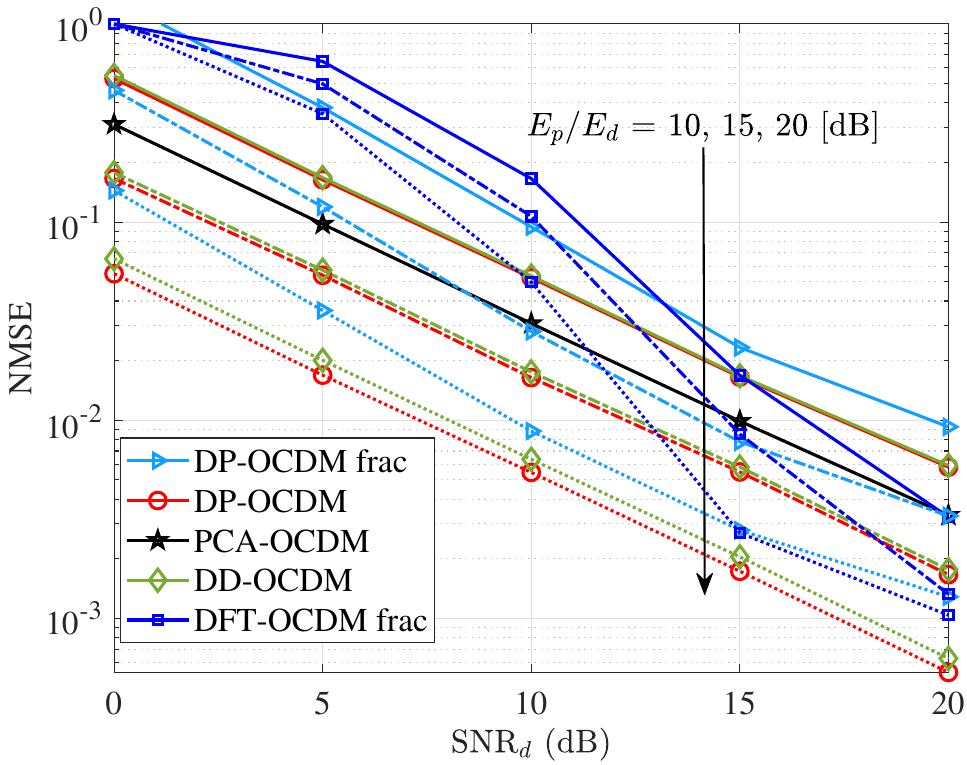}}
\subfloat[BER.\label{fig:ber}]
{\includegraphics[width=0.23\textwidth]{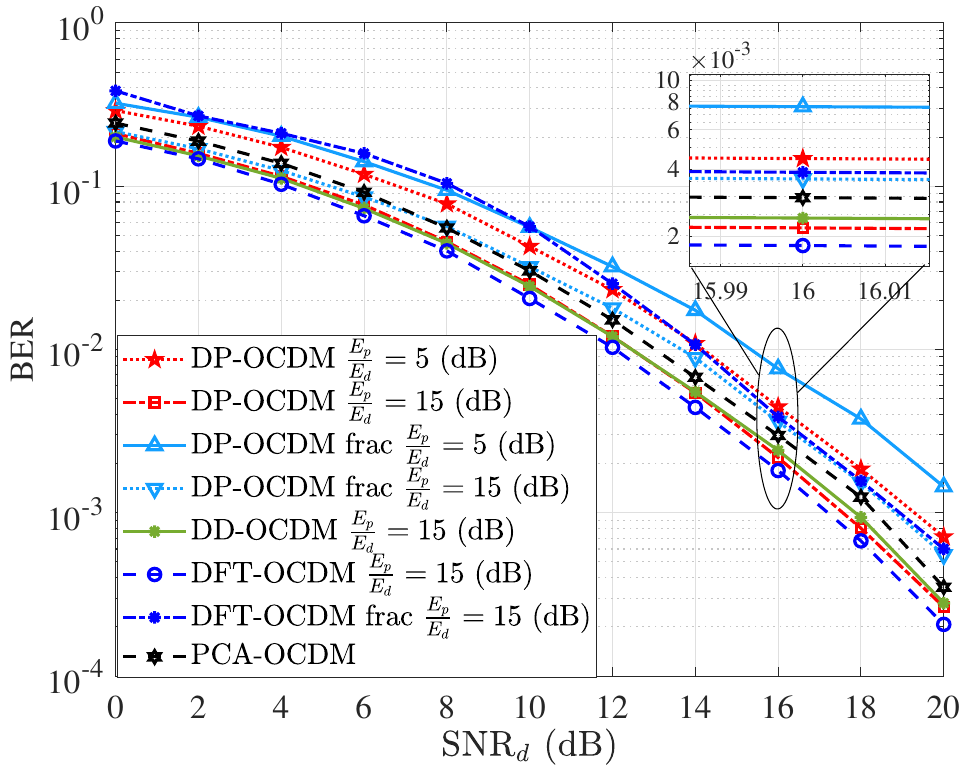}}
\subfloat[CE complexity.\label{fig:complexity}]
{\includegraphics[width=0.23\textwidth]{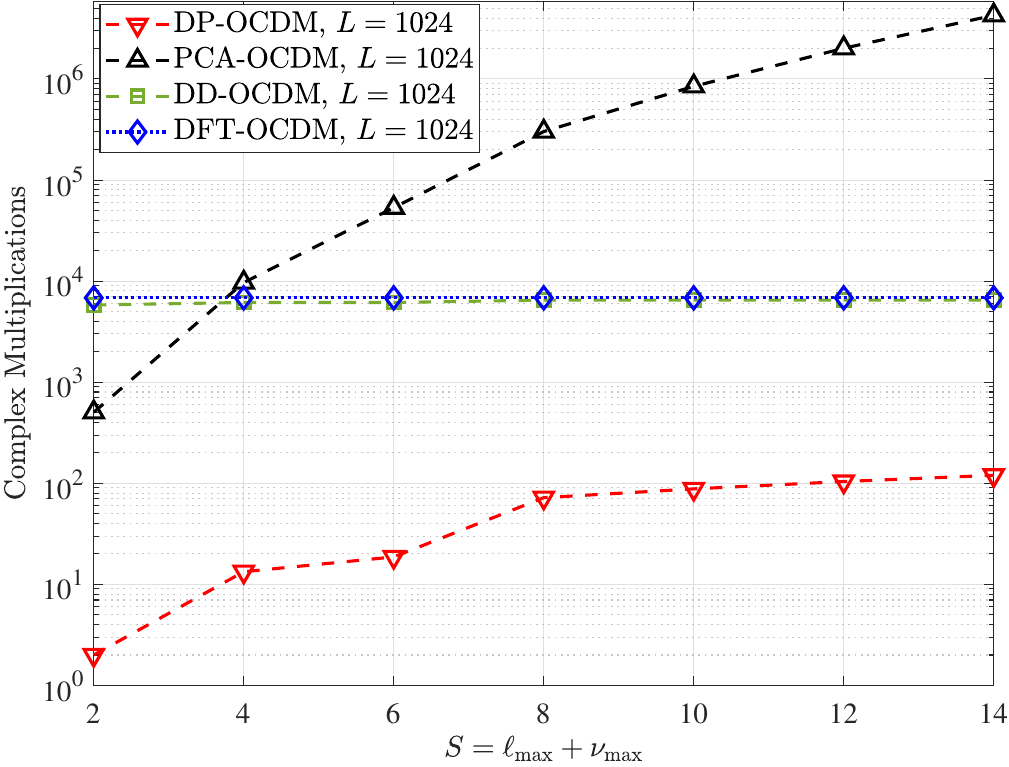}}
\subfloat[Achievable rate.\label{fig:se}]
{\includegraphics[width=0.23\textwidth]{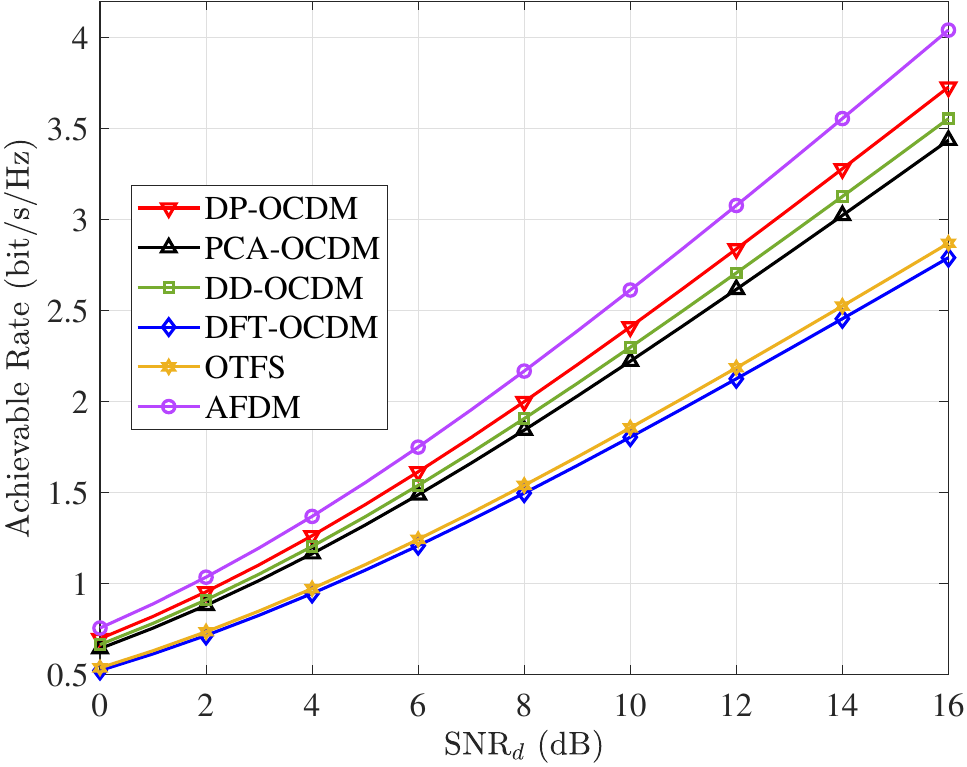}}
\caption{Performance comparison versus $\mathrm{SNR}_d$ in
(a), (b), and (d), and versus $S$ in (c).}
\label{fig:sim_results}
\end{figure*}

Let $\delta_r=|\epsilon_r|$. The offset direction is obtained from
\begin{equation}
s_r=
\begin{cases}
+1, & |z_r(1)|\geq|z_r(-1)|,\\
-1, & \text{otherwise},
\end{cases}
\qquad
\rho_{r,d}=
\frac{|z_r(s_rd)|}{|z_r(0)|},
\label{eq:frac_ratio}
\end{equation}
where $d<M/2$ is a positive integer for which the required diagonal
sample remains available. From the Dirichlet kernel,
\begin{equation}
\rho_{r,d}
=
\frac{\sin(\pi\delta_r/M)}
{\sin\!\left(\pi(d-\delta_r)/M\right)}.
\label{eq:frac_ratio_sine}
\end{equation}
Solving \eqref{eq:frac_ratio_sine} for $\delta_r$ yields
\begin{equation}
\hat{\epsilon}_{r,d}
=
s_r\frac{M}{\pi}
\tan^{-1}\!\left(
\frac{\rho_{r,d}\sin(\pi d/M)}
{1+\rho_{r,d}\cos(\pi d/M)}
\right).
\label{eq:frac_single_est}
\end{equation}
Several valid spacings may be combined as
\begin{equation}
\hat{\epsilon}_r=
\frac{\sum_{d\in\mathcal D_r}
|z_r(s_rd)|^2\hat{\epsilon}_{r,d}}
{\sum_{d\in\mathcal D_r}|z_r(s_rd)|^2},
\qquad
\hat{\widetilde{\nu}}_r=
\hat\nu_{0,r}+\hat{\epsilon}_r ,
\label{eq:frac_final}
\end{equation}
where $\mathcal D_r$ contains the selected positive spacings whose
samples remain inside the available observation region.

Finally, let $\mathcal Q_r$ denote the available diagonal samples used
for coherent gain recovery and define
$\chi_r(u)=e^{\jmath\hat\phi_{k,r}(u)}
D_M(\hat\epsilon_r-u)$. The LS gain estimate is
\begin{equation}
\hat h_r=
\frac{1}{\sqrt{E_p}}
\frac{\sum_{u\in\mathcal Q_r}\chi_r^*(u)z_r(u)}
{\sum_{u\in\mathcal Q_r}|\chi_r(u)|^2}.
\label{eq:frac_gain}
\end{equation}
Thus, fractional-Doppler refinement requires only a few neighboring
samples and scalar operations around each coarse peak, without
iterative search or transceiver modification.

\section{Simulation Results}
\label{sec:simulation}

DP-OCDM is compared with PCA-OCDM~\cite{wang2023pilot},
DD-OCDM~\cite{zegrar2024novel}, and
DFT-OCDM~\cite{haif2024novel}; OTFS~\cite{raviteja2019otfs}
and AFDM~\cite{bemani2023afdm} are additionally included in the
overhead benchmark. Unless otherwise stated, QPSK,
$L=MN=1024$, $N=8$, $M=128$,
$\ell_{\max}=\nu_{\max}=7$, and $N_{\rm CP}=16$ are used, giving
$S=14$ and $K=37$. Relative path powers follow the EVA profile, with
its physical delays mapped to the integer delay grid. The same channel
model and SNR definition are used throughout the comparison, and the
corresponding MMSE data detector is employed for the BER evaluation.
The threshold in \eqref{eq:detection_threshold} uses
$P_{\rm FA}=10^{-3}$. The data SNR is
$\mathrm{SNR}_d=E_d/\sigma^2$, while pilot strength is controlled by
$E_p/E_d$. Integer Dopplers are used in the on-grid case; fractional
offsets satisfy $-1/2<\epsilon_r\leq1/2$, with the path responses kept
sufficiently separated for local refinement. Hence, the
fractional-Doppler results characterize this separated-path regime.

Channel-estimation accuracy is measured by
$$
{\rm NMSE}
=
\mathbb E\!\left[
\frac{\|\widehat{\mathbf H}-\mathbf H\|_F^2}
{\|\mathbf H\|_F^2}
\right],
$$
where $\mathbf H$ and $\widehat{\mathbf H}$ denote the true and
reconstructed channel matrices. Fig.~\ref{fig:nmse} shows that on the
integer grid DP-OCDM localizes each path at the coordinates derived in
Section~\ref{sec:estimation}, yielding competitive estimation accuracy.
Increasing $E_p/E_d$ improves both support detection and gain
estimation. Fractional Doppler spreads each path around its coarse
location and increases NMSE, but the structured diagonal leakage in
\eqref{eq:frac_diag_response} allows the closed-form refinement to
recover much of this loss without windowing.

The uncoded BER in Fig.~\ref{fig:ber} follows the same behavior:
stronger pilots improve channel acquisition and hence detection.
Under fractional Doppler, DP-OCDM remains close to DFT-OCDM at the
higher pilot-energy setting, although DFT-OCDM employs time-domain
windowing to suppress fractional-Doppler sidelobes
\cite{haif2024novel}. DP-OCDM instead estimates the offset directly
from neighboring localized samples.

For Fig.~\ref{fig:complexity}, only integer-grid channel-acquisition
processing additional to conventional OCDM modulation/demodulation is
counted. Let
$S_{\rm PCA}=\ell_{\max}+2\nu_{\max}$ and
$D_{\rm PCA}=(\ell_{\max}+1)(2\nu_{\max}+1)$. The minimum number of
PCA pilot subblocks is~\cite{wang2023pilot}
$$
P_{\min}=
\max\!\left\{
\begin{array}{l}
\min(\ell_{\max},2\nu_{\max})+1,\\[-1mm]
\left\lceil D_{\rm PCA}/(S_{\rm PCA}+1)\right\rceil
\end{array}
\right\}.
$$
For the finite block used in the comparison, $P_{\rm PCA}$ is the
smallest frame-valid $P\geq P_{\min}$ satisfying $P\mid L$ and
$L/P>2S_{\rm PCA}+1$, and
$N_o=P_{\rm PCA}(S_{\rm PCA}+1)$. When the PCA MMSE system is formed
and solved directly, the dominant complexities are
\begin{equation}
\begin{aligned}
C_{\rm DP}
&\simeq\frac{(S+1)N}{3}\log_2N,\\
C_{\rm DFT}
&\simeq\frac{2L}{3}\log_2L,\\
C_{\rm DD}
&\simeq\frac{L}{3}\log_2L+2L
+\frac{L}{3}\log_2N_d,\\
C_{\rm PCA}
&\simeq N_oD_{\rm PCA}^2+D_{\rm PCA}^3 .
\end{aligned}
\label{eq:complexity_all}
\end{equation}
For fixed pilot, channel-statistical, and noise parameters, the PCA
filtering matrix may be precomputed, reducing its online cost; the
curve in Fig.~\ref{fig:complexity} represents direct MMSE-system
formation and solution.

For the complexity sweep,
$\ell_{\max}=\lfloor S/2\rfloor$ and
$\nu_{\max}=S-\ell_{\max}$. At each $S$, DP-OCDM uses the smallest
divisor $N\mid L$ satisfying
$N\geq\max\{\ell_{\max}+1,\nu_{\max}+1\}$ and $L/N>K$.
For DD-OCDM, the smallest valid Zak dimension is selected subject to
$L=M_dN_d=\lambda N_d^2$, $N_d\geq\ell_{\max}+1$, and
$M_d\geq2\nu_{\max}+1$, with integer $\lambda$.
DP-OCDM grows mildly because only $S+1$ short transforms
are evaluated; DFT-OCDM and DD-OCDM are dominated by full-block
transformations, whereas direct PCA MMSE processing grows rapidly with
the BEM dimension. The fractional DP refinement adds only a small
number of per-path neighboring-sample and LS operations and is not
included in this integer-grid complexity comparison.

Fig.~\ref{fig:se} compares an achievable-rate benchmark after
accounting for CP and pilot-guard resources. For
PCA-OCDM,
$\varphi_{\rm PCA}=P_{\rm PCA}(2S_{\rm PCA}+1)$.
For the periodic pilot placement of DD-OCDM, we use the smallest
frame-valid pilot count
$$
N_{p,\rm DD}
=
\min\left\{
d:\,
d\geq\max(\ell_{\max},\nu_{\max}),~
d\mid L
\right\},
$$
with $G_{\rm DD}=2\nu_{\max}+\ell_{\max}$, giving
$\varphi_{\rm DD}=N_{p,\rm DD}(2G_{\rm DD}+1)$
\cite{zegrar2024novel}.

Since the two-pilot DFT-OCDM estimator recovers delay from a phase
relation~\cite{haif2024novel}, several delays sharing one Doppler bin
overlap at the same pilot location. For general multipath, we
therefore adopt a conservative extension, used only for this benchmark,
with $2(\ell_{\max}+1)$ pilot observations, following the standard $2P$
exponential-sample identifiability requirement for
$P=\ell_{\max}+1$ delays, each independently protected against the
relative pilot-data Doppler span $[-2\nu_{\max},2\nu_{\max}]$.

The resulting overheads are
\begin{equation}
\begin{aligned}
\varphi_{\rm DP}
&=NK,\\
\varphi_{\rm PCA}
&=P_{\rm PCA}(2S_{\rm PCA}+1),\\
\varphi_{\rm DD}
&=N_{p,\rm DD}(2G_{\rm DD}+1),\\
\varphi_{\rm DFT}
&=2(\ell_{\max}+1)(4\nu_{\max}+1),\\
\varphi_{\rm OTFS}
&=(2\ell_{\max}+1)(4\nu_{\max}+1),\\
\varphi_{\rm AFDM}
&=2(\ell_{\max}+1)(2\nu_{\max}+1)-1 .
\end{aligned}
\label{eq:overheads}
\end{equation}
For AFDM, the last expression follows from
$c_1=(2\nu_{\max}+1)/(2L)$, which gives
$Q_{\rm AFDM}=2Lc_1\ell_{\max}+2\nu_{\max}$ and
$\varphi_{\rm AFDM}=2Q_{\rm AFDM}+1$
\cite{bemani2023afdm}.

Using a common prefix length and the Shannon expression for the
payload, the achievable-rate benchmark of scheme $b$ is
\begin{equation}
R_b^{\rm ach}
=
\frac{L-\varphi_b}{L+N_{\rm CP}}
\log_2(1+\mathrm{SNR}_d)
\quad\text{bit/s/Hz}.
\label{eq:ach_rate}
\end{equation}
This benchmark isolates the losses due to the prefix and
channel-acquisition overhead while assuming ideal payload detection.

Fig.~\ref{fig:se} shows that AFDM achieves the largest
achievable rate because of its compact affine-domain guard. DP-OCDM follows closely and provides the highest
payload efficiency among the considered OCDM estimators. PCA-OCDM and
DD-OCDM require larger protected regions, while OTFS incurs a
two-dimensional delay-Doppler guard. The conservative generalized
DFT-OCDM benchmark increases once symmetric Doppler support and
repeated-Doppler multipath resolvability are included; its
fractional-Doppler windowing cost is not counted. The result highlights
the key DP-OCDM principle: beyond the gain of modulo-$N$ wrapping,
partitioning the coupled displacement support reduces the shared guard
sufficiently to compensate for the second pulsone and further reduce the total channel estimation overhead.

\section{Conclusion}

This letter proposed DP-OCDM, where two phase-offset
pulsones observe complementary regions of the coupled OCDM
displacement support while sharing a single compact guard. The phase
offset preserves unique pilot association and delay-Doppler recovery
under controlled Doppler wrapping, so that fewer bins per pulsone and
the reduced shared guard jointly lower the overhead. Localized short
transforms recover the channel with very low complexity, and
fractional Doppler is supported through a lightweight refinement.
Simulations show competitive NMSE and BER, with the lowest
channel-acquisition complexity and the highest payload efficiency
among the considered OCDM estimators. Performance bounds and strongly
overlapping fractional delay and Doppler responses are left for future
work.


\end{document}